\documentclass[twoside,a4paper,11pt]{proceedings}
\usepackage{graphicx}
\usepackage{hyperref}
\usepackage{natbib}
\usepackage{amssymb,amsfonts,amsmath,amstext,amsgen,amsopn,amsxtra,indentfirst,lscape,xtab}
\usepackage{setspace}
\newcommand{\arcmin}{$^{\prime}$}
\newcommand{\arcsec}{$^{\prime\prime}$}

\newcommand{\kms}{km\,s$^{-1}$}

\newcommand{\Ha}{H$\alpha$}

\newcommand{\ghafas}{\texttt{GH$\alpha$FaS}}

\begin{document}
\pagenumbering{arabic}
\pagestyle{myheadings}
\thispagestyle{empty}
\vspace*{-1cm}
{\flushleft\includegraphics[width=3cm,viewport=0 -30 200 -20]{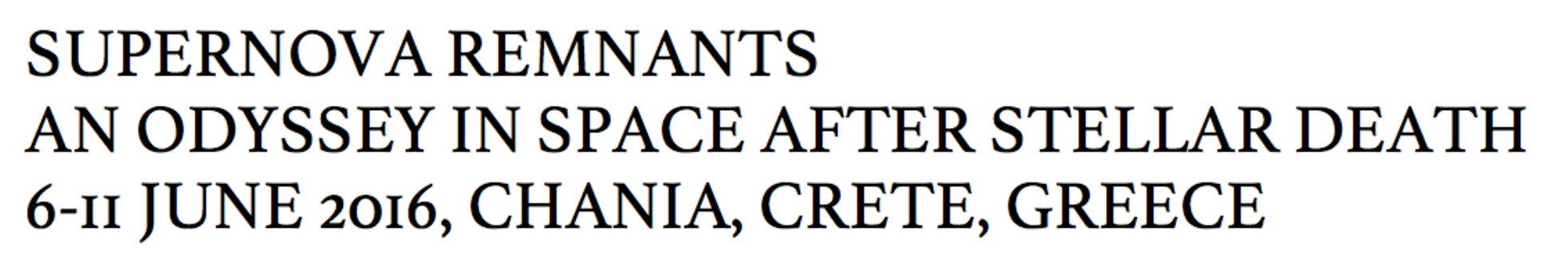}}
\vspace*{0.2cm}
\begin{flushleft}
{\bf {\LARGE
H$\alpha$ Imaging spectroscopy of Balmer-dominated shocks in Tycho's supernova remnant
}\\
\vspace*{1cm}
Sladjana Kne\v{z}evi\'{c}$^1$,
Ronald L\"asker$^2$,
Glenn van de Ven$^3$,
Joan Font$^4$,
John C. Raymond$^5$,
Parviz Ghavamian$^6$,
John Beckman$^4$
%
}\\
\vspace*{0.5cm}
%
\footnotesize{
$^{1}$
Department of Particle Physics and Astrophysics, Faculty of Physics, The Weizmann Institute of Science, Rehovot 76100, Israel \\
$^{2}$
Finnish Centre for Astronomy with ESO (FINCA), University of Turku, V\"ais\"al\"antie 20, FI-21500 Kaarina, Finland \\
$^{3}$
Max Planck Institute for Astronomy, K\"{o}nigstuhl 17, D-69117, Heidelberg, Germany\\
$^{4}$
Instituto de Astrof\'{i}sica de Canarias, V\'{i}a L\'{a}ctea, La Laguna, Tenerife, Spain\\
$^{5}$
Harvard-Smithsonian Center for Astrophysics, 60 Garden Street, Cambridge, MA 02138, U.S.A.\\
$^{6}$
Department of Physics, Astronomy and Geosciences Towson University, Towson, MD 21252, U.S.A.
}
%
\end{flushleft}
\markboth{
Contributed Talk -- Balmer-dominated shocks in Tycho's SNR
}{
Kne\v{z}evi\'{c} et al.
}
\thispagestyle{empty}
%
\begin{minipage}[l]{0.09\textwidth}
\ 
\end{minipage}
\begin{minipage}[r]{0.9\textwidth}
\vspace{0.2in}
\section*{Abstract}{\begin{spacing}{1.0} \small
We present Fabry-P\'erot interferometric observations of the narrow \Ha\ component in the shock front of the historical supernova remnant Tycho (SN 1572). 
Using \ghafas\ (Galaxy \Ha\ Fabry-P\'erot Spectrometer) on the William Herschel Telescope, we observed a great portion of the shock front in the northeastern (NE) 
region of the remnant. The angular resolution of $\sim$1\arcsec\ and spectral resolving power of R$\sim$21\,000 together with the large field-of-view (3.4\arcmin$\times$3.4\arcmin) of the 
instrument allow us to measure the narrow \Ha-line width in 73 bins across individual parts of the shock simultaneously and thereby study the indicators of several 
shock precursors in a large variety of shock front conditions. Compared to previous studies, the detailed spatial resolution of the filament also allows us to mitigate possible artificial 
broadening of the line from unresolved differential motion and projection. Covering one quarter of the remnant's shell, we confirm the broadening of 
the narrow \Ha\ line beyond its intrinsic width of $\sim$20\,\kms\ and report it to extend over most of the filament, not only the previously investigated dense 'knot\,g'. 
Similarly, we confirm and find additional strong evidence for wide-spread intermediate-line ($\sim$150\,\kms) emission. Our Bayesian analysis approach allows us to quantify the evidence 
for this intermediate component as well as a possible split in the narrow line. Suprathermal narrow 
line widths point toward an additional heating mechanism in the form of a cosmic-ray precursor, while  the intermediate component, previously only qualitatively 
reported as a small non-Gaussian contribution to the narrow component, reveals a broad-neutral precursor.
\end{spacing}}
\end{minipage}
\vspace{-0.2in}
\section{Introduction}

The main spectral characteristic of Balmer-dominated shocks (BDSs) is a two-component \Ha\ line. Components are produced in radiative decays of excited 
atoms \citep{ckr80}, whereas cold hydrogen atoms (the ones overrun by the shock) produce a narrow component ($\sim$10\,\kms) and broad-neutrals 
generate a broad component ($\sim$1000\,\kms). Broad-neutrals are formed in a charge exchange (CE) process between a cold atom and a hot proton downstream of the shock.
BDSs are an important diagnostic tool for shock parameters \citep{heng10}: 
narrow (broad) component width indicates the pre (post)-shock temperature; the shock velocity can be estimated from the broad line width and in combination with proper motions of optical filaments provides distance estimates; 
the electron-to-proton temperature ratio can be constrained from the components' widths and their intensity ratios. The \Ha-line profile is also influenced by possible shock precursors (emissions from the shock interacting with the pre-shock medium), for example 
cosmic-rays (CRs) and broad-neutral precursors.  
Heating in the CR precursor results in a narrow \Ha\ component broadened beyond the normal 10--20\,\kms\ gas dispersion \citep{mor13}. Furthermore, CRs transfer also
momentum to the pre-shock neutrals introducing a Doppler shift between the pre- and post-shock gas \citep{lee07}.
CE in the broad-neutral precursor introduces an additional component to the \Ha\ profile -- intermediate component -- with the width of $\sim$150\kms\ \citep{mor12}.

The \Ha-line that is broader than the intrinsic 20\,\kms\ was previously measured in low-spatial resolution data of Tycho \citep{ghava00,lee07}, and interpreted as a strong indicator of CR production. 
The same studies reported also on the detection of the intermediate component. However, both analyses focused on the \Ha-bright, 
but very complex 'knot\,g', where multiple or distorted shock fronts can contribute to the measured broadening of the narrow and intermediate \Ha\ line. 
Just as spatial resolution is crucial to eliminate or reduce the artificial broadening effect of differential projection, extended spatial coverage of the filament can help to ascertain spatially varying shock (and ambient ISM) conditions. For the first time, we provide both of these in our forthcoming analysis. We expand on previous studies also by providing full posterior distributions instead of relying merely on best-fit parameters (line fluxes, centroids and widths), and provide a quantitative assessment of the significance of the intermediate line as well as possible multiple narrow lines (projected shock fronts).

\section{Observations and Reduction}

The instrument setup of \ghafas\ is well suited for our study of spatially and spectrally resolving the narrow \Ha\ line along the Tycho's NE rim.
\ghafas\ is very sensitive to gather enough signal-to-noise (S/N) of around 10 within 6.9\,h of observations, and its field-of-view (FoV) of 3.4\arcmin$\times$3.4\arcmin\ (1024$^2$ pixels at 0.2\arcsec\ pixel scale) is 
large enough to cover roughly one quarter of the whole remnant. The instrument response function
is well approximated by a Gaussian with full width at half maximum (FWHM) of 19\,\kms\ \citep{blasco10}.
The spectral coverage of around 400\,\kms\ was centered around \Ha\ line and split into 48 channels with a sampling velocity resolution of nearly 8\,\kms. 

\begin{figure*}[!t]
\vspace{-0.3in}
\includegraphics[width=0.4\textwidth,angle=-90]{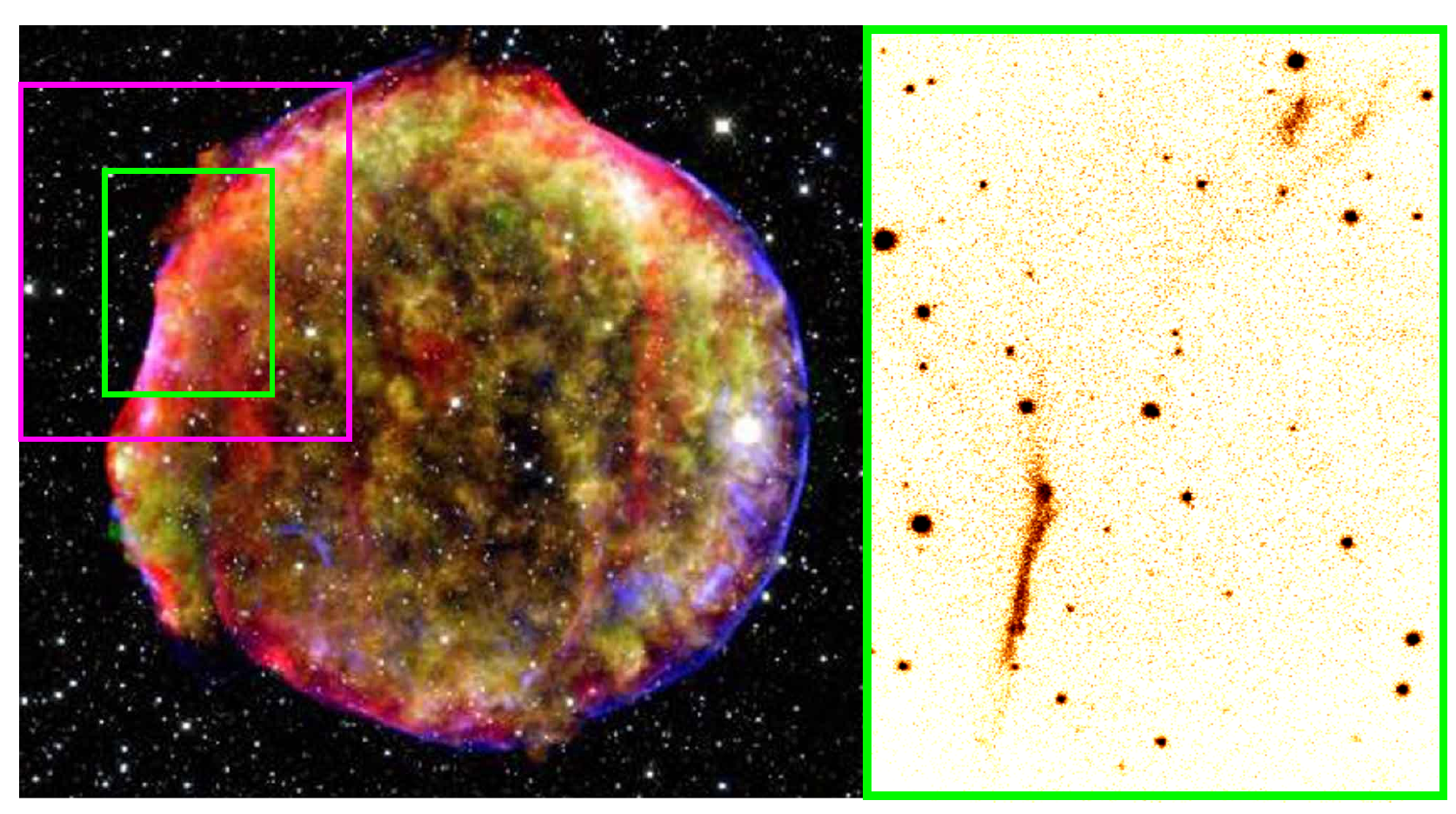} 
\centering
\caption{\footnotesize The left panel shows a composite image of the remnant ($\sim$8\,\arcmin\ in diameter) of Tycho Brahe's 1572 
	supernova, combining data from the Chandra X-ray Observatory (yellow, green, blue; NASA/CXC/SAO), Spitzer Space Telescope 
	(red; NASA/JPL-Caltech), and the Calar Alto Observatory (white stars; Krause et al.). The magenta box indicates 
	the pointing and the FoV of 3\arcmin.4$\times$3\arcmin.4 of the \ghafas\ Fabry-P\'erot interferometer. 
	The green box marks the region which is zoomed-in in the right panel to show our reduced and integrated \ghafas\ \Ha\ image.
	}
	\label{fig:fig1_part1}
\end{figure*}

We reduced the data (see Figure~\ref{fig:fig1_part1}) following the standard procedure for \ghafas\ data described in \citep{hernandez08}. 
For each exposure we performed a phase- and wavelength-calibration and thereby built a data-subcube (x,y,$\lambda$) with 48 calibrated constant-wavelength slices. 
For each frame (exposure, subcube), we then derived a (relative) astrometric solution, and aligned all frames onto a common coordinate system, before stacking. 
In order to be able to take the flatfield effect and the effective exposure time into account, and to reconstruct the local background spectrum, we use the series of object-masked frames to build a flatfield- and a background-stack in the same way and in addition to the "raw" data stack (cube). We then use the flatfield (effective exposure time) cube and the background cube to compare the intrinsic emission model with the data cube (see next section).

\section{Analysis and Results}

\begin{figure*}[!t]
\vspace{-0.2in}
\includegraphics[width=0.4\textwidth,height=0.7\textwidth,angle=-90]{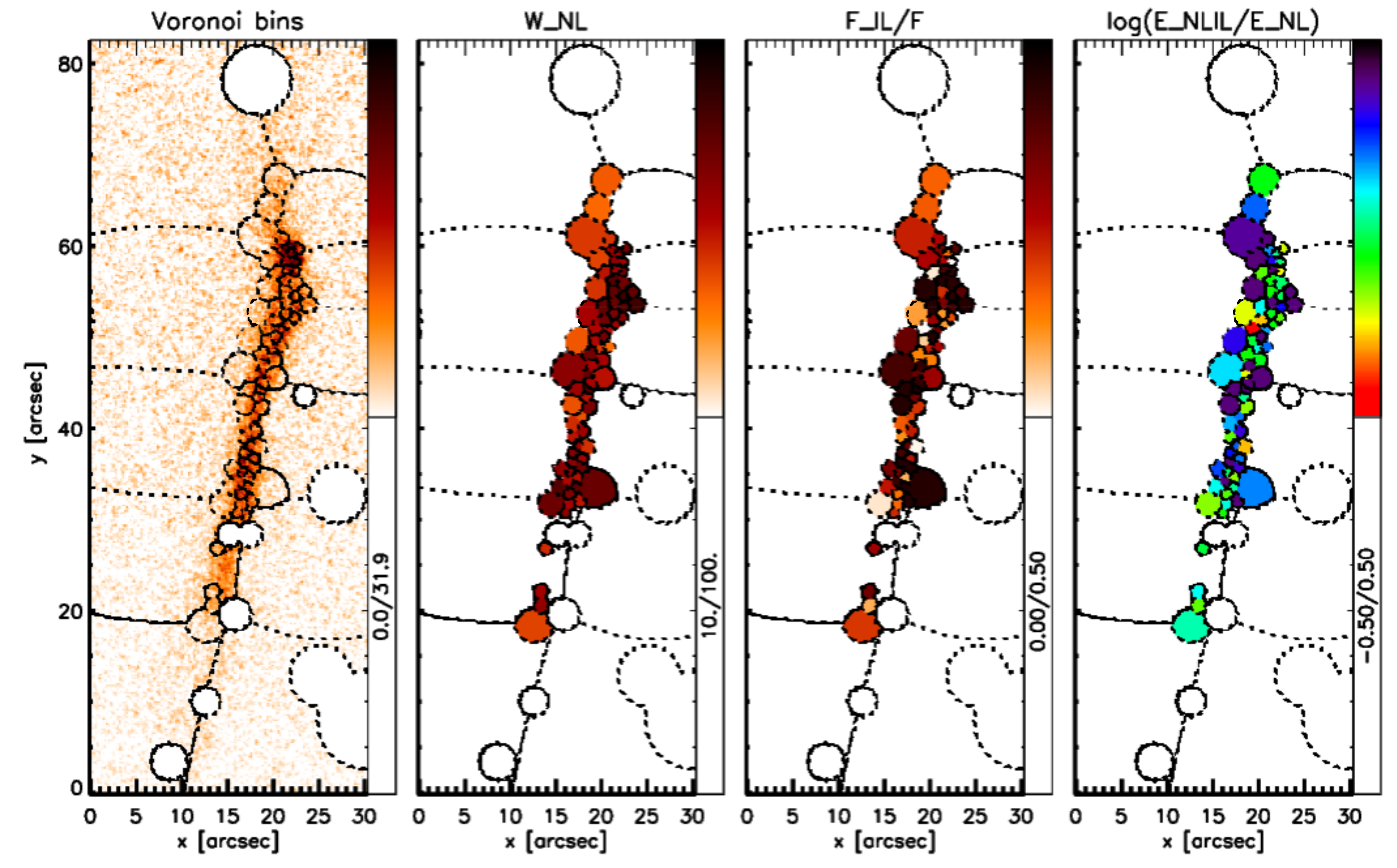} 
\vspace{-0.1in}
\centering
\caption{\footnotesize Spatial maps of Tycho's NE filament: contours of 73 Voronoi bins for S/N=10 binning criteria, spatial variation of the intrinsic narrow-line width ($W_{NL}$) 
	in the NL model, IL flux fraction in the NLIL model, and logarithmic evidence ratio of the NLIL relative to the NL model. 
	The white circles are the masked stars.}
	\label{fig:fig1_part2}
\end{figure*}

Utilizing the large FoV, limiting bin sizes to $>1''$ that corresponds to the seeing conditions, and requiring S/N$\simeq$10 per bin, we were able to extract 
spectra from 73 spatial Voronoi bins \citep{cappellari03}. Thus, we are in a position to measure the narrow \Ha-line widths across individual parts of the 
shocks simultaneously, and thereby study the indicators of precursors in a large variety of shock front conditions (see Figure~\ref{fig:fig1_part2}). 
Although we see only $\sim1\%$ variations between background regions of 25$^2$ pixels, for very large bins with their large integrated signal and signal-to-noise, 
even these small unaccounted-for residuals can become comparable to the noise. Therefore, we exclude bins with an area larger than 400 pixels from the analysis.

We use Bayesian inference to perform parameter estimation -- traditionally termed "fitting" -- as well as comparison and selection. This choice of method is motivated by our desire to maintain small bins with often low S/N, but to simultaneously obtain reliable and complete information about generally significantly non-Gaussian relative probabilities of the parameters ("errors"). This stands in stark contrast to the commonly applied method of deriving some (supposedly) best-fit set of parameters and a rough, generally vastly underestimated, "error bar" via minimizing the chi-square (maximum-likelihood) and calculating its local derivatives. We numerically calculate the posterior parameter probability distributions ("posteriors" for short) via a Markov-Chain Monte-Carlo (MCMC) method, and summarize the posteriors for each bin and model by their global maximum-posterior ("best-fit"), median, and the central 95\%-confidence interval limits. 

For our purpose, just as important as ascertaining the line parameters is the model comparison (and subsequent selection). We want to determine if, as visual inspection of the spectra and theoretical considerations (precursors, differential shock projection) suggest, a so-called intermediate line (IL) is needed in addition to the narrow line (NL) to properly describe the data, or if the NL is split, i.e. if two centroid-shifted NLs are present. Before comparing any model and its specific parameters with the data, we account for the local flatfield spectrum (effective relative exposure time) and add to it the local background spectrum.
Each line (\Ha\ component) is represented by a Gaussian (flux, centroid, width), except for the broad line (BL), which, due to its large width compared to the \ghafas\ spectral range, is represented by a constant that also includes the continuum. We consider the following intrinsic models: NL, 2NL, NLNL, NLIL, 2NLIL, NLNLIL, where 2NL and NLNL denote two NLs with common and independent width, respectively.

\begin{figure*}[!t]
\vspace{-0.25in}
\centering
\includegraphics[width=0.7\textwidth,height=0.9\textwidth, angle=-90]{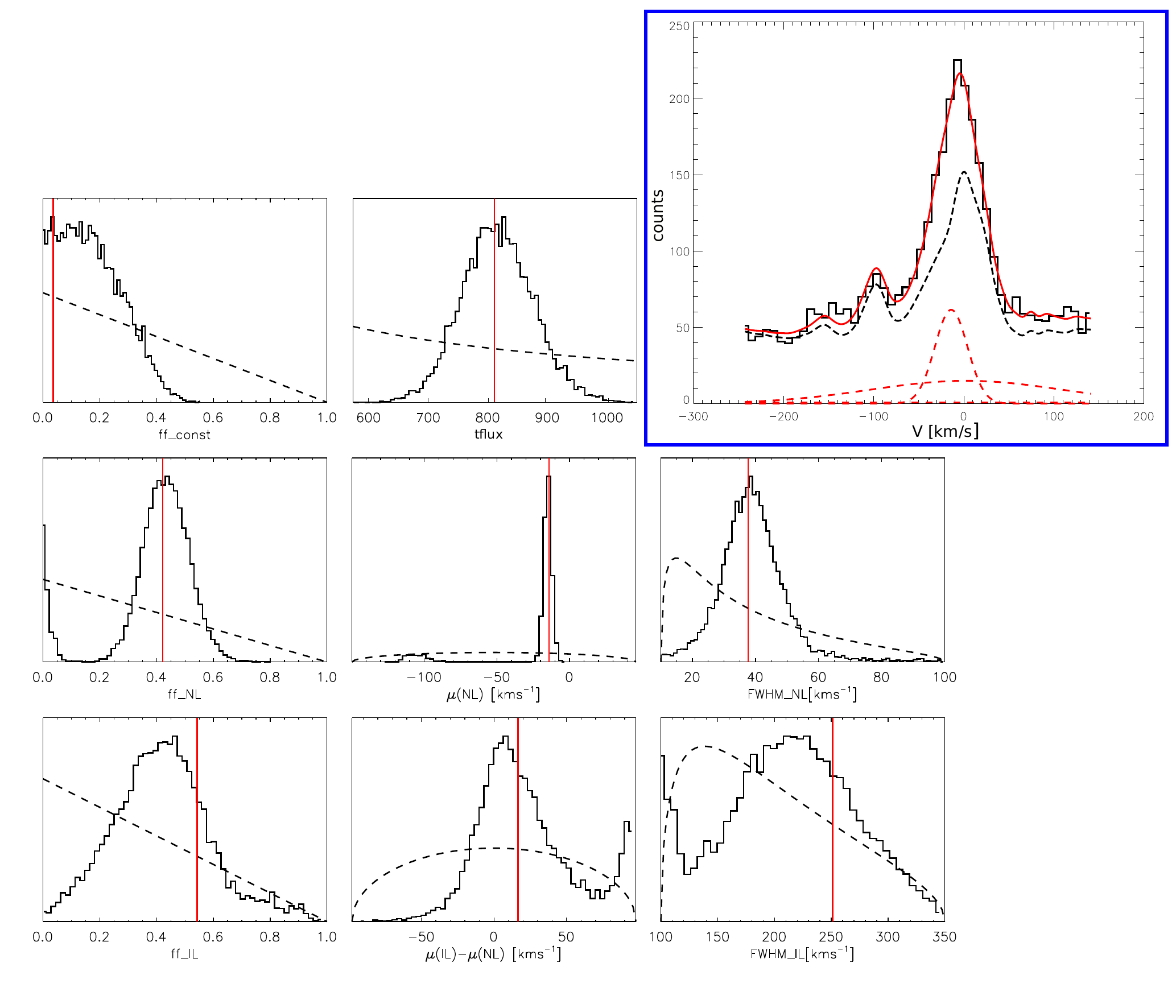} 
\vspace{-0.25in}
\caption{\footnotesize The top-right blue panel shows the spectrum of one of the bins (solid-black line), the background model (dashed-black line) and components of the intrinsic ``best-fit'' model (dashed-red lines).
                The ``best-fit'' model is overplotted with the solid red line.
                The remaining 8 panels are 1D-marginalized posteriors over model parameters (solid-black lines): total intrinsic flux, flux fractions in the continuum and lines, NL centroid, IL offset 
                from the NL centroid, and line widths. Dashed-black lines are prior distributions and vertical red lines are the estimated parameters of the ``best-fit'' model, i.e. the parameters at the 
                maximum of the posterior distribution.}
	\label{fig:fig2}
\end{figure*}

The prior parameter distributions are flat or nearly so (Beta or Dirichlet distributions with low $\alpha(=\beta) \geq 1$ parameter) -- that is, we do not (strongly) prefer any parameter values within the model definition limits. Those limits are [10, 100]\,\kms\ for the NL width (intrinsic FWHM), and for the IL, [100, 350]\,\kms\ as predicted for shock velocities [1500, 3500]\,\kms\ \citep{mor12,mor13}.
Figure~\ref{fig:fig2} shows 1D-marginalized posteriors over model NLIL parameters for one of the bins. The vertical red lines show the parameters at the maximum of the posterior 
(``best-fit'' model, the prediction of which is the solid red line in the top-right blue-framed panel). We find the NL width being close to 40\,\kms\ and IL with the width of $\approx$\,250\,\kms\ comprising 55\% 
of the total flux. Moreover, NL widths in a single-line model (NL model) are larger than 20\,\kms\ (the second panel in Figure~\ref{fig:fig1_part2}), that is they are broadened beyond the expectation when no CR precursor is present. This result is not entirely new \citep{ghava00,lee07}, but we now and here show it to not be an artifact of spatially averaging over regions with potentially differential motion along the line-of-sight. 

In addition, we see that a significant fraction of the bins also show prominent IL (third panel in Figure~\ref{fig:fig1_part2}) in the NLIL model. 
Here, as a preliminary and "pedestrian" indicator, this statement is underpinned by IL-NL flux fractions; the stronger the best-fit IL line the more "significant" it is. However, a more statistically well-defined statement on the need for an IL is based on the models' Bayesian Evidence and their ratios ("Bayes factors"). We use the leave-one-out cross validation (LOO-CV) method to calculate the evidence \citep{bjc12}. It is a derivative version of the standard direct numeric integration (marginalization) of the posterior, with the advantage that the prior enters the result only to second order. 
Figure~\ref{fig:fig2} shows the example where the NLIL model is favoured, for which we get that the logarithmic evidence ratio of the NLIL relative to the NL model is about 1\,dex. This means that an IL in addition to a single NL explains our data 10 times better than a simple NL model, irrespective of any particular parameter values -- notably, also not only considering the respective best-fit parameters. The logarithmic evidence ratio of the NLIL relative to the NL model 
for other Voronoi bins is shown in the Figure~\ref{fig:fig1_part2}.

\begin{figure*}[!t]
\vspace{-0.35in}
\centering
\includegraphics[width=0.9\textwidth,height=0.5\textwidth]{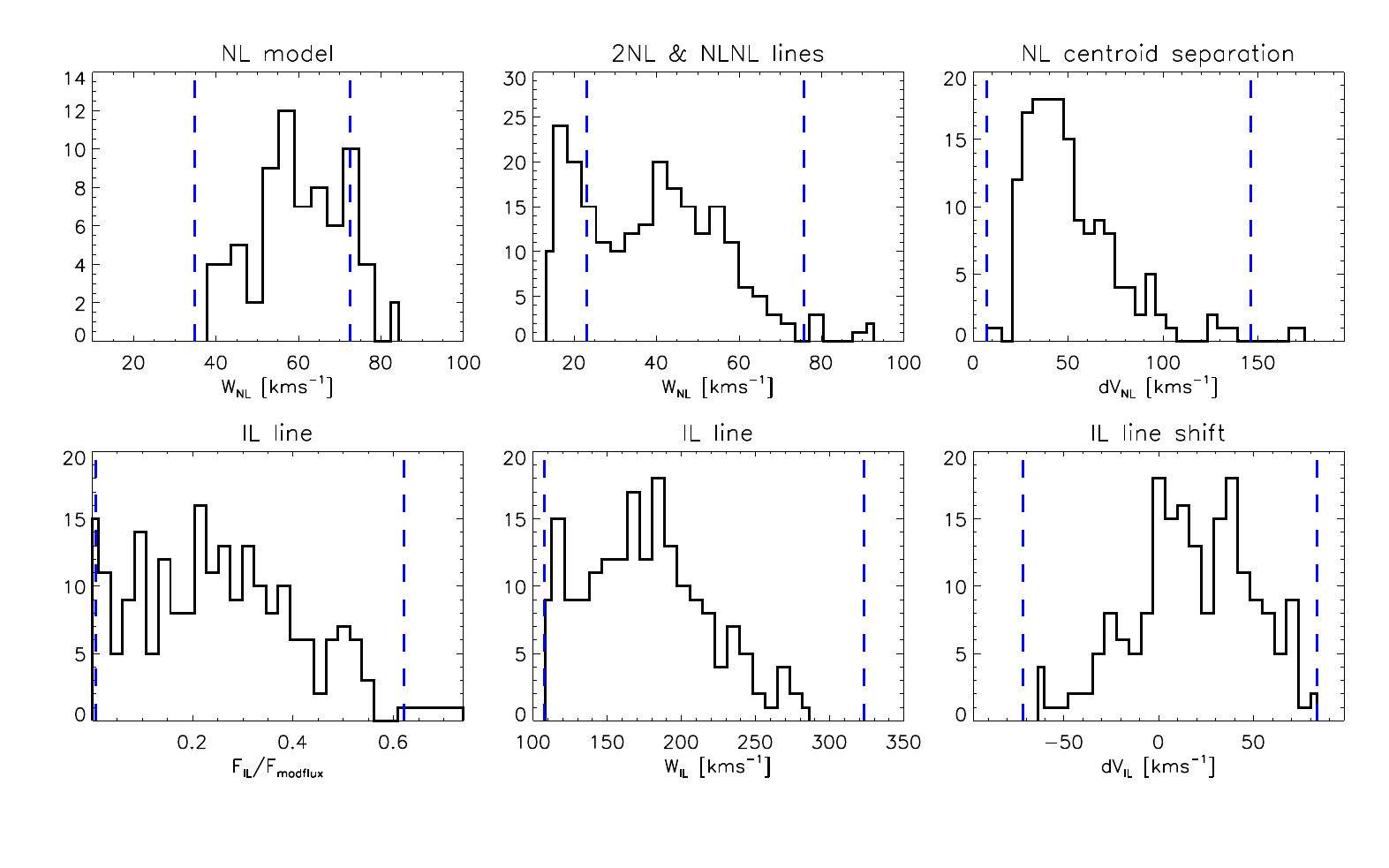} 
\vspace{-0.4in}
\caption{\footnotesize Summarized results for all Voronoi bins. Top panels: ``best-fit'' intrinsic NL widths in NL model, NL widths in 2NL(IL) and NLNL(IL) models, NL centroid separation in all double-NL models.
                Bottom panels: IL flux fractions, IL widths, IL offsets from the NL centroid in all IL models. Vertical dashed-blue lines show the mean of the 2.5\% (left line) and 97.5\% (right line) quantile distributions.}
	\label{fig:fig3}
\end{figure*}

In Figure~\ref{fig:fig3} we summarize our results for all Voronoi bins and all 
the models regardless of the favoured ones. We plot histograms of the best-fit intrinsic NL widths in the NL model, NL widths and their centroid separation in 2NL(IL) and NLNL(IL) models,  
IL flux fractions, widths and their offsets from the NL centroids. For each of the estimated parameter we 
also calculate central 95\%-confidence interval. The vertical blue-dashed lines show the mean of the 2.5\% (left line) and 97.5\% (right line) quantile distributions. 
We often find NL widths being much larger than 20\,\kms: $\approx$\,60\,\kms\ is the mean of the NL width in single-NL models; even in the 2NL(IL) and NLNL(IL) models, it is generally $\approx$40\,\kms. 
The intrinsic IL widths are $\approx$\,180\,\kms\ on average with 30\% flux contribution to the total intrinsic flux. 

\section{Summary}

We presented the narrow \Ha\ spectroscopic observations of Tycho's NE Balmer filaments. This study provides spectroscopic data that is for the first time spatially resolved (spectro-imagery), 
with large coverage that comprises and resolves the entire NE filament.
Our analysis includes Bayesian model comparison that enables a quantitative, probabilistic and well-defined model comparison. 
We find that the broadening of the NL beyond 20\,\kms\ that was noted in previous studies was not an artifact of the spatial integration, and that it extends across the whole filament, not only the previously covered 'knot\,g'. 
Likewise, we confirm the suspected presence of an IL, and show it to be widespread.
The first result points toward the evidence of heating in the CR precursor, while the second result reveals the presence of the broad-neutral precursor. 

%
\section*{Acknowledgments}   
%
We would like to thank Coryn Bailer-Jones (MPIA) for useful discussion on Bayesian statistics and Giovanni Morlino (INFN Gran Sasso Science Institute) for discussion on 
theoretical predictions for the shock emission in Tycho's SNR.

\setlength{\bibsep}{0.0pt}
\footnotesize
\bibliographystyle{aj}
\bibliography{proceedings}

\end{document}